\documentclass[
  aps,
  prb,
  twocolumn,
  superscriptaddress,
  amsmath,
  amssymb,
  floatfix,
  nofootinbib,
  longbibliography
]{revtex4-2}

\usepackage{graphicx}
\usepackage{siunitx}
\usepackage{xcolor}
\usepackage[colorlinks=true,allcolors=blue]{hyperref}
\usepackage{hyperref}
\graphicspath{{figures/}}

\newcommand{\GN}{G_\mathrm{N}}
\newcommand{\GS}{G_\mathrm{S}}

\newcommand{\Vfg}{V_\mathrm{FG}}
\newcommand{\Vsd}{V_\mathrm{SD}}
\newcommand{\Bperp}{B_\perp}
\newcommand{\Bpar}{B_\parallel}
\newcommand{\dIdV}{dI/dV_\mathrm{SD}}
\newcommand{\Dind}{\Delta_\mathrm{ind}}

\begin{document}

\title{Andreev-Enhanced Conductance in a Bilayer Graphene Quantum Point Contact}

\author{Emma Horgan}
\affiliation{Faculty of Physics, University of Vienna, Boltzmanngasse 5, 1090 Vienna, Austria}
\affiliation{Vienna Doctoral School in Physics, University of Vienna, Boltzmanngasse 5, 1090, Vienna, Austria}

\author{Bernd Aichner}
\affiliation{Faculty of Physics, University of Vienna, Boltzmanngasse 5, 1090 Vienna, Austria}

\author{Nadine Omar}
\affiliation{Faculty of Physics, University of Vienna, Boltzmanngasse 5, 1090 Vienna, Austria}

\author{Kenji Watanabe}
\affiliation{Research Center for Electronic and Optical Materials,
             National Institute for Materials Science,
             1-1 Namiki, Tsukuba 305-0044, Japan}

\author{Takashi Taniguchi}
\affiliation{Research Center for Materials Nanoarchitectonics,
             National Institute for Materials Science,
             1-1 Namiki, Tsukuba 305-0044, Japan}

\author{Luca Banszerus}
\email{luca.banszerus@univie.ac.at}
\affiliation{Faculty of Physics, University of Vienna, Boltzmanngasse 5, 1090 Vienna, Austria}
             
\date{\today}

\begin{abstract}  \noindent Quantum point contacts provide a platform for probing mode-dependent Andreev reflection. Here, we demonstrate Andreev-enhanced conductance in a gate-defined bilayer graphene quantum point contact coupled to a superconducting aluminum electrode. Electrostatic gating enables sequential mode occupation, verified by normal-state conductance quantization. In the superconducting state, the conductance is enhanced within the conductance plateaus, while the enhancement decreases between plateaus, consistent with the nonlinear transmission dependence of Andreev reflection. The enhancement yields effective transmission probabilities up to 0.9, and extends over a bias window of approximately 140 $\mu$V, which we associate with a proximity-induced superconducting gap, close to the gap of the aluminum film. For in-plane fields, the enhancement persists up to the critical field, and the bias window follows the spectral gap predicted by Abrikosov--Gor'kov theory. In contrast, perpendicular fields suppress the enhancement approximately linearly from zero field, deviating from homogeneous pair breaking and indicative of flux penetration.
\end{abstract}

\maketitle


At the interface between a normal conductor and a superconductor, charge transport at energies within the superconducting gap is mediated by Andreev reflection. In this process, an incident electron is reflected as a hole while a Cooper pair is transferred into the superconductor~\cite{Andreev1964}. The probability of this process depends nonlinearly on the transmission of the normal conductor: While normal-state conductance scales linearly with transmission, Andreev reflection requires both the electron and the reflected hole to transmit~\cite{BTK1982,Beenakker1992}. Consequently, superconductivity can enhance the conductance of highly transmitting channels by up to a factor of two, while suppressing that of poorly transmitting channels. 

\noindent Semiconductor-superconductor hybrid devices provide a versatile platform for investigating Andreev reflection and proximity-induced superconductivity. Early experiments demonstrated gate-controlled Andreev reflection in semiconductor quantum point contacts (QPCs)~\cite{Takayanagi1995Sep} and quantized supercurrents through superconducting constrictions~\cite{Takayanagi1995Nov,Irie2014Apr}. More recently, highly transparent semiconductor-superconductor interfaces have enabled a wide range of experiments, including conductance doubling and hard induced gaps in InAs QPCs~\cite{Kjaergaard2016} and Andreev spectroscopy in hybrid nanowires~\cite{Doh2008Dec,Lee2014Jan}. Gate-tunable Josephson transport in InAs and Ge heterostructures has further enabled superconducting diodes~\cite{Baumgartner2022Jan,Ciaccia2023Aug,Valentini2024Jan}, unconventional current-phase relations~\cite{Spanton2017Dec,Ueda2020Sep,Ciaccia2024Jan}, and hybrid superconducting qubits~\cite{Larsen2015Sep,Sagi2024Jul}. Most recently, QPCs in Ge/SiGe heterostructures exhibited Andreev-enhanced conductance quantization consistent with effective interface transmissions approaching 0.9~\cite{Kiyooka2026Sep}. Together, these advances establish gate-defined semiconductor nanostructures as sensitive probes of superconducting proximity effects and motivate extending this approach to materials with additional electronic degrees of freedom.\\
Graphene offers a promising platform for semiconductor-superconductor hybrid devices, combining high electronic mobility with spin and valley degrees of freedom. Bilayer graphene (BLG) adds an electrically tunable band gap~\cite{McCann2006Gap,Zhang2009}, which allows nanostructures to be defined electrostatically rather than by etching~\cite{Overweg2018Jan,Eich2018Jul}. This has enabled QPCs with conductance quantization in steps of $4e^2/h$~\cite{Overweg2018Jan,Overweg2018Dec,Kraft2018,Lee2020Mar} and quantum dots reaching the single-carrier regime~\cite{Eich2018Jul,Moller2021Dec,Banszerus2023Jun}, with the fourfold spin and valley degeneracy weakly lifted by Kane-Mele spin-orbit coupling~\cite{Banszerus2020May,Banszerus2021SpinValley,Kurzmann2021Kondo}. Graphene-based hybrid devices have demonstrated gate-tunable and ballistic Josephson transport~\cite{Heersche2007Mar,Calado2015Sep,BenShalom2016Apr}, specular Andreev reflection, phase-dependent Andreev bound states~\cite{Efetov2016,Bretheau2017}, and gate control of the Andreev spectrum in BLG Josephson junctions~\cite{Park2024ABS}. Very recently, Andreev-enhanced conductance quantization has been demonstrated in a gate-defined BLG quantum point contact~\cite{GalanteAgero2026} and proximity-induced superconductivity has been probed by gate-defined quantum dots~\cite{Schmidt2026}. 
In this Letter, we investigate mode-dependent Andreev reflection and proximity-induced superconductivity in a gate-defined BLG QPC coupled to an aluminum electrode. Comparing the normal and superconducting conductance, we observe a conductance enhancement on quantized plateaus that decreases between plateaus, consistent with the nonlinear transmission dependence of Andreev reflection. The enhancement yields effective channel transmissions of up to 0.9 and extends over a bias window corresponding to a proximity-induced superconducting gap of approximately $\Delta_\mathrm{ind} = \SI{140}{\micro\electronvolt}$. Magnetic-field-dependent measurements reveal qualitatively different behavior in the two field orientations. For in-plane fields, the enhancement persists nearly up to the critical field, while its bias window follows the spectral gap predicted by Abrikosov--Gor'kov pair-breaking theory~\cite{AbrikosovGorkov1961,Skalski1964} rather than the superconducting order parameter. In contrast, perpendicular fields suppress the enhancement from zero field onward, which we attribute to flux penetration into the aluminum film.


\begin{figure}
  \includegraphics[width=\linewidth]{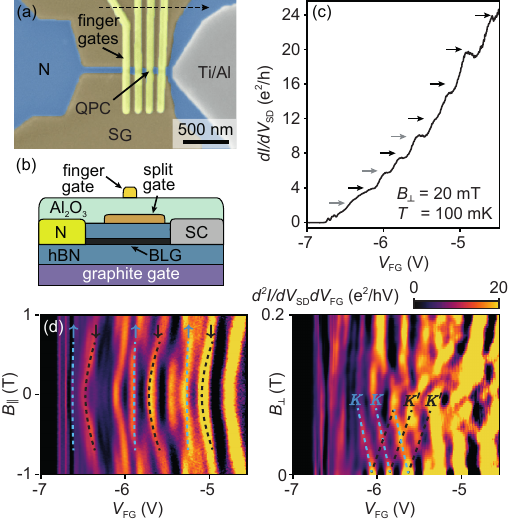}
  \caption{
    (a) False-color scanning electron micrograph of the normal-metal-QPC-superconductor (N-QPC-S) device.
    (b) Schematic cross section of the van der Waals heterostructure and gate stack along the dashed arrow in (a), illustrating the graphite back gate, split gates (SGs), and finger gates (FGs).
    (c) Normal-state differential conductance $dI/dV_\mathrm{SD}$ as a function of finger-gate voltage $V_\mathrm{FG}$, exhibiting quantized plateaus at multiples of $4e^2/h$ (black arrows) and additional plateaus at odd multiples of $2e^2/h$ (grey arrows).
    (d) Transconductance $d^2I/dV_\mathrm{SD}dV_\mathrm{FG}$ as a function of $V_\mathrm{FG}$ and in-plane magnetic field $B_\parallel$, showing the evolution of the conductance plateaus.
    (e) Same as (d) for a perpendicular magnetic field $B_\perp$, revealing a valley splitting of the subbands.}
  \label{fig1}
\end{figure}

Figure~\ref{fig1}a shows a scanning electron micrograph of our normal-metal--QPC--superconductor (N-QPC-S) device, which is based on a van der Waals heterostructure comprising a BLG flake encapsulated between 59~nm and 70~nm thick hexagonal boron nitride (hBN) flakes and placed on a graphite flake serving as a back gate. The heterostructure was assembled using a dry stacking technique. Normal-metal (N) Cr/Au (5/80~nm) and superconducting (S) Ti/Al (5/100~nm) electrodes were fabricated by reactive ion etching followed by electron-beam evaporation. Cr/Au split gates (SGs), separated by a 100~nm wide gap and positioned 100~nm from the superconducting contact, provide the lateral confinement for the QPC. A set of 70~nm wide Cr/Au finger gates (FGs), electrically isolated from the SGs by a 20~nm thick Al$_2$O$_3$ layer, provides electrostatic control along the channel. A schematic cross section of the heterostructure and gate stack is shown in Figure~\ref{fig1}b.

Transport measurements were performed in a dilution refrigerator using four-terminal DC and low-frequency lock-in techniques. To define the QPC, we apply a back-gate voltage of $V_\mathrm{BG}=8$~V and a split-gate voltage of $V_\mathrm{SG}=-6$~V. The resulting displacement field opens a band gap in the BLG regions beneath the split gates, where the Fermi energy is tuned into the gap, rendering these regions insulating. In contrast, the region between the split gates remains conducting, forming a narrow, highly doped channel. Applying a negative voltage to the finger gate indicated in Figure~\ref{fig1}a locally depletes the channel and controls the number of transmitted modes. 

We first characterize the QPC in the normal state at $T=100$~mK and $\Bperp=20$~mT, above the critical field of the aluminum contact. Figure~\ref{fig1}c shows the differential conductance, $\dIdV$, as a function of finger-gate voltage, $\Vfg$. The conductance exhibits plateaus at multiples of $4e^2/h$, consistent with the fourfold spin and valley degeneracy of BLG. At low mode numbers, additional plateaus appear at odd multiples of $2e^2/h$, indicating a partial lifting of the fourfold degeneracy.

To elucidate the origin of the additional conductance steps, we follow the evolution of these modes in an in-plane magnetic field (see Figure~\ref{fig1}d). The subband features associated with the additional $2e^2/h$ plateaus remain distinguishable up to $B_\parallel=1$ T and exhibit an increasing, hyperbolic separation with magnetic field (see dashed lines in Figure~\ref{fig1}d). We do not resolve additional steps at multiples of $e^2/h$. We interpret this as a signature of the Kane-Mele spin-orbit coupling, which splits the fourfold degenerate subbands into the two Kramers doublets $(K\!\uparrow, K'\!\downarrow)$ and $(K\!\downarrow, K'\!\uparrow)$, separated by the Kane-Mele spin-orbit gap, $\Delta_\mathrm{SO}\approx 60~\mu$eV~\cite{Banszerus2020May}. An in-plane field couples only to the spin, so that both Kramers pairs are expected to remain twofold degenerate with a separation $\sqrt{\Delta_\mathrm{SO}^2 + (g\mu_\mathrm{B}B_\parallel)^2}$. A valley splitting induced by the confinement would instead leave each level spin degenerate. Figure~\ref{fig1}e shows the evolution in a perpendicular magnetic field, where the Berry-curvature-induced orbital magnetic moment gives rise to a valley splitting, resulting in a considerably stronger, linear magnetic-field dependence (see dashed lines in Figure~\ref{fig1}e)~\cite{Lee2020Mar, Banszerus2020May}. We observe a small residual conductance that remains when the QPC is pinched off, which we attribute to parallel conduction paths beneath the split-gated region. All conductance data are corrected for this parallel contribution and the series resistance of the BLG channel, as described in detail in the Supporting Information (see Supplementary Figure~S1).

\begin{figure}
  \includegraphics[width=\linewidth]{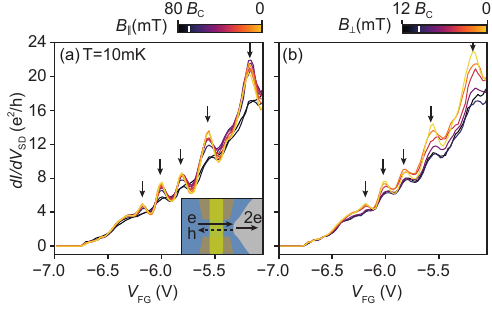}
  \caption{(a) Differential conductance, $dI/dV_\mathrm{SD}$, as a function of finger gate voltage $V_\mathrm{FG}$ for in-plane magnetic fields $B_\parallel$ ranging from zero to 80~mT, exceeding the critical field $B_\mathrm{c,\parallel}$ of the superconducting Ti/Al contact. Inset: Schematic illustration of Andreev reflection, where an incident electron is retroreflected as a hole while a Cooper pair with charge $2e$ is transferred into the superconducting contact. (b) Same measurement for perpendicular magnetic fields $B_\perp$ ranging from zero to 12~mT, exceeding the corresponding critical field $B_\mathrm{c,\perp}$. Both orientations show an enhancement of the conductance in the superconducting state, most pronounced on the quantized conductance plateaus (see arrows).
 }
  \label{fig2}
\end{figure}

We now turn to the superconducting transport properties of the QPC by reducing the temperature to $T=10$~mK and setting the magnetic field to zero. At energies below the superconducting gap, transport is mediated by Andreev reflection, where an incident electron is retroreflected as a hole while a Cooper pair enters the superconductor (inset of Figure~\ref{fig2}a). For a fully transmitting channel, this process doubles the conductance compared to the normal state. At finite transmission, the enhancement is reduced because both the incident electron and the reflected hole must pass through the constriction. For a single channel with transmission probability $\tau$, the ratio between the superconducting ($\GS$) and normal-state ($\GN$) conductance at zero bias and temperature is given by~\cite{Beenakker1992}
\begin{equation}
    R = \frac{\GS}{\GN} = \frac{2\tau}{(2-\tau)^2},
    \label{eq:ratio}
\end{equation}
which reaches $R=2$ for perfect transmission and falls below unity for sufficiently low transmission.

Figures~\ref{fig2}a and \ref{fig2}b show the differential conductance, $\dIdV$, as a function of finger-gate voltage for a series of in-plane and out-of-plane magnetic fields, respectively. At zero magnetic field, the conductance is enhanced compared to the normal state, particularly on the quantized conductance plateaus. Increasing the magnetic field progressively suppresses this enhancement until the normal-state conductance is recovered above the critical field of the superconducting contact. The enhancement vanishes at critical fields of $B_\mathrm{c,\parallel}=72$~mT and $B_\mathrm{c,\perp}=9$~mT, corresponding to an anisotropy of approximately eight. This strong anisotropy is consistent with the geometry of the superconducting thin film. Using $B_\mathrm{c,\perp}=\Phi_0/(2\pi\xi^2)$, we estimate a superconducting coherence length of $\xi\approx190$~nm. The thin-film relation $B_\mathrm{c,\parallel}/B_\mathrm{c,\perp}=\sqrt{12}\,\xi/d$ yields an estimated anisotropy of approximately six for the film thickness $d=105$~nm, in reasonable agreement with the experiment. The aluminum contact has a critical temperature of $T_\mathrm{c}=1.02$~K (see Supplementary Figure~S2), slightly below the bulk value, consistent with a weak suppression by the thin titanium adhesion layer. 

\begin{figure}
  \includegraphics[width=\linewidth]{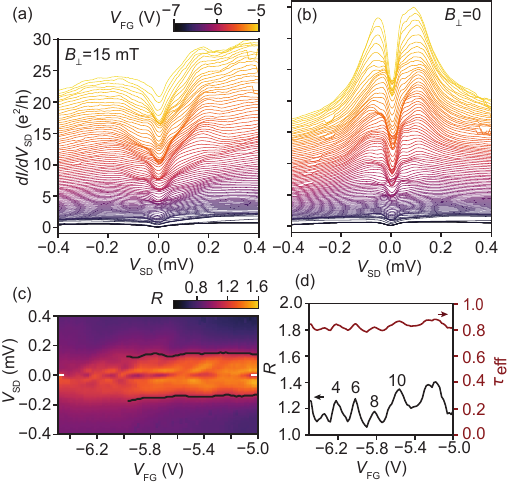}
  \caption{(a) Waterfall plot of the differential conductance, $dI/dV_\mathrm{SD}$, as a function of source-drain bias $\Vsd$ for a series of finger gate voltages $\Vfg$, with
  the aluminum contact driven normal by a perpendicular magnetic field of
  $\Bperp = 15$ mT. (b) Same as (a) at $\Bperp = 0$, with the aluminum contact
  superconducting. (c) Ratio of the superconducting and normal differential conductance, $R$, as a function of $\Vsd$ and $\Vfg$. (d) $R$ at $\Vsd = 0$ (see white markers in c) (left axis) and the effective transmission $\tau_\mathrm{eff}$ extracted from Eq.~\eqref{eq:ratio} (right axis) as a function of $\Vfg$. Numbers indicate the conductance of the corresponding
  plateaus in units of $e^2/h$. }
  \label{fig3}
\end{figure}

Interestingly, the superconducting conductance exhibits pronounced oscillations across the conductance steps, with maxima on the quantized plateaus and minima between them, where it approaches the normal-state conductance (see arrows in Figure~\ref{fig2}). Similar features have been observed in semiconductor-superconductor QPCs~\cite{Kjaergaard2016} and predicted theoretically for coherent ballistic constrictions~\cite{Beenakker1992,Mortensen1999}. The nonlinear dependence of Andreev reflection on the transmission eigenvalues makes the superconducting conductance particularly sensitive to variations in individual channel transmissions. In addition, a reduction of the enhancement below the ideal factor of two has been predicted for point contacts supporting only a few propagating modes, with a pronounced dependence on the position of the Fermi level. In particular, coherent scattering and resonances associated with the opening of additional subbands can give rise to pronounced conductance oscillations even in the absence of disorder~\cite{Mortensen1999}.


To investigate the bias dependence of the Andreev enhancement, we measure the differential conductance as a function of source-drain voltage, $V_\mathrm{SD}$, for different finger-gate voltages, $V_\mathrm{FG}$. Figure~\ref{fig3}a shows the normal-state conductance recorded at $B_\perp=15$~mT, revealing a pronounced suppression around zero bias. Interestingly, this suppression is strongly gate-dependent, being most prominent on the quantized conductance plateaus and considerably weaker between them. With increasing $|V_\mathrm{SD}|$, the conductance generally increases, and additional plateau-like features emerge at finite bias, suggesting contributions from higher-energy transport modes. The low-bias suppression occurs on a characteristic bias scale corresponding to approximately $50~\mu$eV, comparable to the Kane-Mele spin-orbit splitting reported in BLG nanostructures~\cite{Banszerus2020May}. This suggests a possible connection to the additional $2e^2/h$ conductance plateaus observed in Figure~\ref{fig1}c. In the superconducting state (Figure~\ref{fig3}b), a similar overall bias dependence is observed, together with a pronounced conductance enhancement around $|V_\mathrm{SD}| < 140\, \mu$V.

To isolate the contribution from Andreev reflection, we calculate the ratio $R(V_\mathrm{SD},V_\mathrm{FG})=G_\mathrm{S}/G_\mathrm{N}$ between the superconducting and normal-state differential conductance at the same source-drain bias and finger-gate voltage (Figure~\ref{fig3}c).
From the characteristic width of the conductance-enhancement window, within which $R>1.1$, we estimate an effective superconducting gap scale of approximately $140\,\mu$eV (see black lines in Figure~\ref{fig3}c), which we associate with proximity-induced superconductivity in the BLG region. This value is slightly below the gap of the aluminum film, $\Delta_\mathrm{Al}\approx155~\mu$eV, obtained from its measured critical temperature of $T_\mathrm{c}=1.02$~K. The approximately gate-independent energy scale suggests that the enhancement is primarily governed by the superconducting contact rather than the gate-dependent subband structure of the QPC. The conductance enhancement is maximal at zero bias and decreases towards the edges of the bias window, rather than peaking at the gap edge as expected for a tunnel barrier at the interface from Blonder-Tinkham-Klapwijk theory. A similar behavior has been reported for quantum point contacts in Ge/SiGe and InAs heterostructures before and was attributed to the induced gap of the proximitized region~\cite{Kiyooka2026Sep,Kjaergaard2016}.

\begin{figure}
  \includegraphics[width=\linewidth]{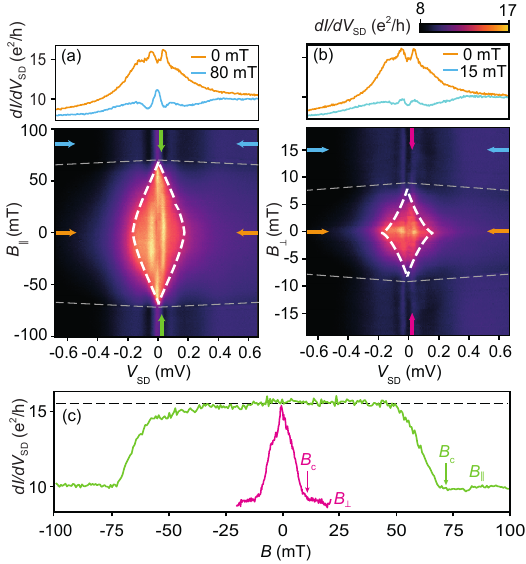}
  \caption{Differential conductance, $dI/d\Vsd$, as a function of $\Vsd$ and
  (a) in-plane and (b) out-of-plane magnetic field. The line cuts above each map compare the zero-field conductance (orange) with the normal-state conductance measured above the respective critical field (blue). White dashed lines indicate the spectral gap, $e\Vsd=\pm\Omega_\mathrm{g}(B)$,  calculated from homogeneous Abrikosov--Gor'kov pair-breaking theory using $\alpha\propto\Bpar^2$ in (a) and $\alpha\propto|\Bperp|$ in (b). The grey dashed lines indicate the switching of the aluminum contact. (c) Conductance at a fixed source-drain bias indicated in (a) and (b), as a function of magnetic field, for both field orientations.}
  \label{fig4}
\end{figure}

Figure~\ref{fig3}d shows the zero-bias conductance ratio, $R(V_\mathrm{FG})$, exhibiting pronounced oscillations correlated with the quantized conductance plateaus. The enhancement reaches approximately $R=1.4$ on the plateaus and is strongly reduced between them, approaching unity in some regions. Using the single-channel relation in Eq.~\eqref{eq:ratio}, we convert the measured ratio into an effective transmission probability $\tau_\mathrm{eff}$, shown on the right axis of Figure~\ref{fig3}d. The resulting values reach approximately 0.9, indicating highly transmitting channels. Given the few-mode nature of the constriction, $\tau_\mathrm{eff}$ should be understood as a combined measure of the constriction and the interface rather than as the transmission of the constriction alone. This is close to the mode-independent transmission coefficient of $0.88$ and the corresponding ratio of $1.4$ reported for Andreev-enhanced conductance quantization in Ge/SiGe quantum point contacts~\cite{Kiyooka2026Sep}.


Next, we investigate the magnetic-field dependence of the Andreev-enhanced conductance. Figures~\ref{fig4}a and \ref{fig4}b show the differential conductance of the $G_\mathrm{N}$=10~$e^2/h$ plateau as a function of source-drain bias and magnetic field, applied parallel and perpendicular to the sample plane, respectively. The line cuts above each map compare the zero-field conductance (orange lines) with that measured above the critical fields (blue lines), illustrating the suppression of the enhancement in the normal state. In both field orientations, the bias window of enhanced conductance narrows with increasing magnetic field, reflecting the suppression of proximity-induced superconductivity.

To describe this evolution, we calculate the spectral gap $\Omega_\mathrm{g}(B)$ using Abrikosov--Gor'kov pair-breaking theory~\cite{AbrikosovGorkov1961,Skalski1964}, shown as white dashed lines in Figures~\ref{fig4}a,b. Under pair breaking, the order parameter $\Delta$ and the spectral gap $\Omega_\mathrm{g}$ are no longer equal. For a pair-breaking energy $\alpha$ they are described by $\ln(\Dind/\Delta)=(\pi/4)(\alpha/\Delta)$ and $\Omega_\mathrm{g}=\Delta[1-(\alpha/\Delta)^{2/3}]^{3/2}$, where $\Dind=140$~$\mu$eV is the zero-field induced gap determined in Figure~\ref{fig3}c. At finite field, with increasing $\alpha$ the spectral gap closes before the order parameter does, opening a bias range $\Omega_\mathrm{g}<e\Vsd<\Delta$ in which pairing persists but quasiparticle states are already available. For parallel fields we assume the quadratic orbital pair breaking of a thin film, $\alpha\propto\Bpar^2$~\cite{Maki1964}, and for perpendicular fields a homogeneous model with $\alpha\propto|\Bperp|$. For each field orientation, the pair-breaking strength is normalized to the independently measured critical field of the aluminum contact, $B_{\mathrm{c},\parallel}=72$~mT and $B_{\mathrm{c},\perp}=9$~mT, using the Abrikosov--Gor'kov critical pair-breaking condition. Assuming that the induced gap follows the spectral gap of the aluminum film, both curves contain no additional fitting parameters. Importantly, for in-plane magnetic fields, the observed boundaries of the Andreev-enhanced conductance closely follow the calculated spectral gap $\Omega_\mathrm{g}(B)$ rather than the superconducting order parameter $\Delta(B)$, see also Supplementary Figure~S3. This supports our interpretation of the enhancement window as a measure of the proximity-induced spectral gap in BLG. At biases above $\Omega_\mathrm{g}/e$ the order parameter is still finite, but quasiparticle states are available and the Andreev reflection probability falls rapidly, such that the conductance enhancement is suppressed well before $e\Vsd$ reaches $\Delta$. For perpendicular magnetic fields the response is qualitatively different. The conductance enhancement is suppressed from zero field onward, and a characteristic energy scale can no longer be resolved above $\approx 0.7\,B_{\mathrm{c},\perp}$. This suggests additional mechanisms beyond homogeneous pair breaking. One possibility is spatially inhomogeneous superconductivity in the aluminum film, where vortices or normal regions introduce low-energy quasiparticle states~\cite{Abrikosov1957,Hess1989}. Alternatively, orbital effects in the proximitized bilayer graphene could dephase the electron and the retroreflected hole. This mechanism requires a flux of order $\Phi_0/2$ through the area enclosed by the two trajectories, so that the observed field scale of about $7$~mT would correspond to a region of roughly $380\times380$~nm$^2$. This greatly exceeds the distance of about $100$~nm between the split gates and the superconducting contact, which disfavors the graphene region as the dominant mechanism, although diffusive trajectories may enclose a larger area than the geometric separation suggests. We therefore consider flux penetration into the aluminum film the more likely origin. A qualitatively similar behavior is observed on different conductance plateaus and summarized in the Supporting Information (see Supplementary Figure~S4).

To quantify the field dependence of the enhancement, we extract line cuts at a fixed, low but finite source-drain bias, as indicated in Figures~\ref{fig4}a and \ref{fig4}b. The resulting traces are shown in Figure~\ref{fig4}c. We avoid zero bias because of the previously discussed conductance dip associated with the bias dependence of the QPC. For parallel fields, the conductance remains enhanced and essentially constant almost up to $B_{\mathrm{c},\parallel}$ and then rapidly decreases to the normal conductance. The nearly constant enhancement in parallel fields is consistent with robust subgap Andreev reflection and a mostly field-independent normal-state transmission. Within homogeneous pair-breaking theory the Andreev reflection probability deep inside the spectral gap does not depend on $\alpha$, so that no reduction is expected as long as a gap remains open. In contrast, for perpendicular fields, the enhancement decreases approximately linearly from zero field. The absence of any plateau suggests that perpendicular fields suppress Andreev reflection through additional orbital effects beyond the homogeneous pair-breaking description.


In summary, we have investigated Andreev-enhanced conductance in a gate-defined bilayer graphene quantum point contact coupled to a superconducting aluminum electrode. The superconducting conductance is enhanced on the quantized plateaus and the enhancement decreases between them, reflecting the nonlinear dependence of Andreev reflection on the transmission. From the conductance enhancement we extract effective channel transmissions of up to 0.9, while the bias window indicates a proximity-induced superconducting gap of approximately $140~\mu$eV. In parallel magnetic fields, the enhancement remains nearly constant up to the critical field, while its bias window follows the spectral gap predicted by Abrikosov--Gor'kov pair-breaking theory. In contrast, perpendicular magnetic fields suppress the enhancement approximately linearly from zero field, which we attribute to flux penetration into the aluminum film. Our results establish gate-defined BLG quantum point contacts as probes of Andreev transport and proximity-induced superconductivity, providing a foundation for future studies of Andreev bound states and nonlocal superconducting correlations~\cite{Park2024ABS,Schmidt2026,Tan2015CPS}.

\begin{acknowledgments}
This research was supported by the Scientific Service Units of ISTA through access to the Nanofabrication Facility. We particularly thank Juan Luis Aguilera and Nasima Afsharimani for their assistance in developing fabrication recipes. We also thank Stephan Puchegger and the Faculty Center for Nano Structure Research at the University of Vienna for support with scanning electron microscopy. E.H.\ acknowledges financial support from the Vienna Doctoral School in Physics at the University of Vienna. K.W.\ and T.T.\ acknowledge support from the JSPS KAKENHI (Grant Numbers 21H05233 and 23H02052), the CREST (JPMJCR24A5), JST and World Premier International Research Center Initiative (WPI), MEXT, Japan for the growth of hexagonal boron nitride crystals.
\end{acknowledgments}

\bibliographystyle{apsrev4-2}
\bibliography{references}

\end{document}